\documentclass[runningheads]{llncs}
\usepackage[T1]{fontenc}
\usepackage{graphicx}
\usepackage{url}
\newtheorem{mydef}{Definition}
\usepackage{comment}
\usepackage{tcolorbox}

\begin{document}

\title{Towards a Formalisation of Value-based Actions and Consequentialist Ethics}
%

%
\titlerunning{Value-based Actions and Consequentialist Ethics}  
%
\author{Adam Wyner\inst{1}\orcidID{0000-0002-2958-3428} \and Tomasz Zurek\inst{2}\orcidID{0000-0002-9129-3157} \and
Dorota Stachura-Zurek\inst{3}\orcidID{0000-0002-8157-4932}}
\authorrunning{Adam Wyner et al.} 
%
\tocauthor{Adam Wyner, Tomasz Zurek, Dorota Stachura-Zurek}
\institute{
Department of Computer Science, Swansea University, Swansea, UK\\
\email{a.z.wyner@swansea.ac.uk},
\and
Complex Cyber Infrastructure, Informatics Institute, Faculty of Science, University of Amsterdam,\\
\email{t.a.zurek@uva.nl},
\and
Independent researcher,\\
\email{dorota.stachura1@gmail.com}}
\maketitle              

%
%
%
%
%
\begin{abstract}
Agents act to bring about a state of the world that is more compatible with their personal or institutional values.
To formalise this intuition, the paper proposes an action framework based on the STRIPS formalisation. Technically, the contribution expresses actions in terms of Value-based Formal Reasoning (VFR), which provides a set of propositions derived from an Agent's value profile and the Agent's assessment of propositions with respect to the profile.
Conceptually, the contribution provides a computational framework for a form of consequentialist ethics which is \textit{satisficing}, \textit{pluralistic}, \textit{act-based}, and \textit{preferential}.
\end{abstract}

\section{Introduction}

In a multi-agent system, the behaviour of individual agents may be guided by a variety of factors, e.g., in a Belief-Desire-Intention model \cite{bratman1987}, propositions model the state of the world, the goals, and the actions to change states. However, it leaves open by what means an agent selects the propositions to reason about, as the agent selects some propositions that are used to represent a partial state of the beliefs about the world and a partial goal state. To make such a selection, we consider \textit{motivated reasoning} \cite{Kunda1990}, which aims to address how an agent selects propositions relative to the agent's values; the selected propositions can then be used in the Agent's knowledge base from which actions are constructed. In this way, the research addresses ethical agents in the sense that it provides a foundational, computational analysis of the underlying motivations for an agent's behaviour, which can be further developed to address matters of coordination and interaction to represent organisational/institutional behaviour.


What is it for an Agent to behave with respect to their personal or institutional values? There are a range of values. Consider personal values such as self-enhancement (ambition, wealth), openness to change (freedom, daring), conservation (obedience, national security) amongst others \footnote{\cite{Schwartz2012} and values in the EU's Joint Research Centre\\ \url{https://op.europa.eu/webpub/jrc/jrc-values-identities/what-are-values.html}}. Individuals do specific actions which reflect their value preferences: if one highly values wealth, then one may practice frugality and seek high paying employment.
Moreover, some values may relate to ethical (or moral) choices in terms of how individuals and institutions ought to behave (or not) towards others and the world they live in, e.g., self-transcendence (social justice and protecting the environment), though others might be less directly linked to ethical or moral preferences, e.g. self-enhancement. As a caveat, the aim of this paper is not to propose a \textit{moral machine}, one which optimises an agent's actions with respect to the ``best'' moral position. Rather, we provide a means to define acceptable actions with respect to an agent and their values.
In this paper, we take an abstract position towards such relations or sets of values in order to provide a generic framework about how actions reflect value preferences.

We assume that an Agent performs actions in order to bring about a situation (state of the world) that is more compatible with that Agent's values than it had been prior to the execution of the action; that is, the Agent performs some action to realise their values. 
While an Agent may have other reasons to act, here we focus on values. While values are abstract, they can be realised or instantiated via the performance of an action which results in a world that is more compatible with their values than it was before the performance of the the action.
The proposal is grounded in Value-based Formal Reasoning (VFR) (presented in \cite{wynerzurek2023} and section \ref{sec:linkingAgentsPropositionsValues}), wherein the locus of values is based on each Agent’s value profile and how they regard a proposition in view of their values. From this, we construct an Agent’s base of acceptable (from a values point of view) propositions. 
The VFR provides a framework for the attribution of values to propositions, on which we base an action language in the 
familiar state transition mechanism of STRIPS - STRIPS$_{VFR}$ (Section \ref{sec:STRIPS}). Thus, the actions satisfy the Agent's value preferences in terms of bringing about a state more compatible with their value profile. Note that this is distinct from attributing the value to the transition per se (Section \ref{sec:RelatedWork}).
Moreover, since our model focuses on consequences of actions rather than actions itself, it can be used as a representation of kind of \textit{consequentialist ethics}, which is discussed below.


The key novel technical contribution is to use the VFR (Section \ref{sec:linkingAgentsPropositionsValues}) to define STRIPS$_{VFR}$ (Section \ref{sec:STRIPS}), which is an action language, where Agents execute actions to bring about states of the world relative to their value preferences. The key novel conceptual contribution is to provide consequentialist ethics with a formal framework in STRIPS$_{VFR}$ (Section \ref{sec:Consequentialism}). We also introduce a proof-of-concept, an experimental implementation of our model. The paper closes with related work, discussion, and conclusion.

\section{Consequentialism}\label{sec:Consequentialism}

\textit{Consequentialism} stands for theories which assume that normative properties depend on consequences \cite{anscombe1958,sep-consequentialism}, thus representing one of the major approaches to normative ethics.
The foundation for modern consequentialism was laid by Jeremy Bentham, the creator of utilitarianism, for whom the goodness of actions is judged by whether they maximize utility, understood as the greatest happiness for the greatest number of people \cite{Bentham1780-BENITT}. One should therefore follow a course of action which promotes pleasure (happiness) and avoid it if causes pain (unhappiness). This qualitative approach naturally raised serious reservations, paving way to the development of Bentham's ideas and provoking the ongoing discussion on qualitative differences between pleasures, the pluralism of values constituting happiness, or approaches to assessing consequences.
An interesting form of consequentialism relevant to research here was formed in response to one of the strongest objections against consequentialism, the \textit{demandingness} objection. Considering that maximizing consequentialist approaches which require that we maximize good and act in such a way as to bring about the best possible consequences, many critics find them too demanding or objectionable on other grounds \cite{Vallentyne2006-VALAMA-2}. \textit{Satisficing consequentialism} holds that it is morally permissible to choose consequences which are less than optimal, but good enough \cite{SlotePettit1984,McKay2022}.

In this paper, STRIPS$_{VFR}$ seems to most closely model consequentialism which is \textit{satisficing}, \textit{pluralistic} (i.e., a variety of irreducible values), \textit{act-based} (i.e., consequences are evaluated for a single act), and \textit{preferential} (i.e., the agent seeks their own value preference fulfillment).
Finally, we are not advocating that this form of consequentialism is the only or best form of ethical reasoning, but only that it is most compatible with the formalism provided here, which then provides the means to examine the form and implications computationally.

\section{Agents, Propositions, Values, and Weights} \label{sec:linkingAgentsPropositionsValues}

In this section, we outline a value-based language for decision making:

\begin{itemize}
    \item \texttt{Agent}, where each element is an agentive entity.
    \item \texttt{Prop}, where each element is a proposition\footnote{The proposal is neutral as to whether these are atomic or complex, provide a semantic representation or are simply strings.}
    \item \texttt{IncompProp} is of type \texttt{Prop} $\times$ \texttt{Prop}, where for every pair $<$x,y$>$ of distinct elements of type \texttt{Prop},  x and y cannot co-occur in any state. The relation is symmetric. The expressions of such a pair are called \textit{objectively incompatible}, otherwise, they are objectively compatible.
    \item \texttt{Value}, where each element of \texttt{Value} is an abstract object that expresses a value concept such as \textit{freedom}, \textit{security}, etc.
    \item \texttt{Scale}, which is a totally ordered, finite set of scalar elements.
    \item \texttt{Weight} is of type \texttt{Scale} $\vert$ ?. \footnote{$\vert$ is Backus-Naur Form for `or'.} ? is a designated entity type which indicates that a weight is indeterminate (not relevant). While there may be alternative interpretations of `weights', here they reflect the relative `importance' to an agent, e.g., \textit{family} might be a very important value and \textit{personal status} very unimportant.
    As ? is unordered with respect to other elements of \texttt{Scale}, any expression comparing ? to the other entity is false.
\end{itemize}

We can quantify over any type in the basic vocabulary with variables and constants of each type, represented by Greek and Latin subscripts respectively, e.g., for \texttt{Agent}, \texttt{agent$_{\alpha}$}, \texttt{agent$_{\beta}$}, $\ldots$ or \texttt{agent$_{a}$}, \texttt{agent$_{b}$}, $\ldots$.


We first construct an agent's \textit{value profile}, \texttt{AgentValueToWeight}, which indicates the degree of importance that the agent ascribes to a value, where the higher the weight, the more important and the lower the weight the less important. 
\texttt{AgentValueToWeight} is a total function.

\begin{itemize}
\item [] \texttt{AgentValueToWeight} =\\ (\texttt{Agent} $\times$ \texttt{Value}) $\rightarrow$ \texttt{Weight} 
\end{itemize}
We indicate each agent's value profile with a subscript, e.g., \texttt{AgentValueTo\-Weight}$_{agent_{k}}$ for agent agent$_{k}$.
Given the $?$ weight, the importance an agent associates with a value can be indeterminate. 
Since differences in \texttt{AgentValueToWeight} of particular agents represents differences in the levels of importance the agents puts on those values, they can be seen as a subjective or personal value profiles of those agents.

To represent how an agent assesses an element of type \texttt{Prop} with respect to an element of type \texttt{Value} and an element of type \texttt{Weight}, we introduce a total function: 
\begin{itemize}

\item [] \texttt{AgentValuePropWeight}:\\ (\texttt{Agent} $\times$ \texttt{Value} $\times$ \texttt{Prop})  $\rightarrow$ \texttt{Weight}
\end{itemize}
This expresses an agent's disposition towards a proposition with respect to values and weights. While \texttt{AgentValuePropWeight} is a total function, the use of ? signals that there may be propositions which are not meaningfully assessed.
The association of a proposition with a particular value and weight appears in \cite{zurekgvr} and in work on case-based reasoning with factors and values \cite{capon-wyner2013}.

The functions \texttt{AgentValueToWeight} and \texttt{AgentValueProp\-Weight} are taken to be conceptually distinct. In particular, we interpret \texttt{AgentValueTo\-Weight} to indicate the importance that the agent ascribes to the Value, where the higher the weight, the more important, and the lower the weight, the less important, relevant, or disposed. It is a predicate that expresses a disposition of the agent in general about a Value. In contrast, we interpret \texttt{AgentValueProp\-Weight} to indicate an agent's assessment of the \texttt{Value} and \texttt{Weight} ascribed to a proposition.
In other words, \texttt{AgentValueToWeight} expresses an agent's ``ideal'' and ``global'' view on \texttt{Value}s, while \texttt{AgentValuePropWeight} expresses an agent's assessment of each proposition with respect to a value. In this approach, values are directly associated with propositions relative to agents and their value profile.

To reflect an agent's value-based \textit{world view},
we gather all the propositions that are in some sense ``compatible'' with an agent's values. In VFR, this means that the weight an agent assigns to a proposition relative to a value must not be less than the weight that agent assigns to that value in general. The proposition must ``pass'' the filter of acceptability relative to the agent's value profile\footnote{Note that propositions can be indeterminate (represented by ?) with respect to the Weight on Values; that is, not every proposition need be value-weighted. Given ? is unordered with respect to weights, such a proposition always passes an agent's filter}. By ``passing'' a filter, we understand that the weight the agent assigns to a particular proposition in the light of particular value is no less than the weight the agent assigns to this value (see def. \ref{def:propBaseClean}).

Now we abstract from consideration of Values and Weights and extract those propositions which represent the agent's value-based \textit{world view}.
A set \texttt{propBaseClean$_{agent_h}$} contains all and only those propositions which pass the value-weight filter of agent$_h$ for all values. It is important to empahsise that this is a strict, formally convenient definition, which assesses all propositions and creates a set of only those propositions that pass the agent's value filter. We acknowledge matters are more complex and less strict; in other work, we develop more lenient, flexible definitions, though space precludes their presentation here.

\begin{mydef}[propBaseClean]\label{def:propBaseClean}
\texttt{propBaseClean} is of type  \texttt{Prop}.\\
Where $agent_\alpha$ is a variable for elements of type \texttt{Agent}, $p_\beta$ is a variable for elements of type \texttt{Prop}, and $v_\gamma$ is a variable for elements of type \texttt{Value}, the denotation relative to an $agent_\alpha$ is:\\
\texttt{propBaseClean}$_{agent_\alpha}$ = $\{p_\beta | \neg ($\texttt{AgentValuePropWeight}$(agent_\alpha, v_\gamma, p_\beta)  < \\ $\texttt{AgentValueToWeight}$(agent_\alpha, v_\gamma)) \} $
\end{mydef}


Values and weights discriminate amongst propositions. For a particular agent, a lower weight on a particular value implies that there is a lower discriminatory threshold on the acceptability of propositions, which themselves are associated with that Value and the Weight. Simply put, if an Agent has a lower Weight on a particular Value, then more propositions may pass the filter, as they have higher weights on the same value. 
The higher the weight means that an agent has higher standards with respect to the value; there is greater discrimination such that fewer propositions pass the  filter. An intuitive example may help to clarify the relations between the value-weights in AgentValueToWeight and AgentValuePropWeight. Suppose an agent is not much bothered by the quality of coffee, so on the value \textit{taste} the weight is low, and the agent acts accordingly. When they are served a coffee that on the value \textit{taste} has weight that is also low, then the agent drinks the coffee; when served another coffee with value taste with weight high, then the agent also drinks the coffee. In effect, the taste makes little difference to this agent, as they don't discriminate. On the other hand, suppose the agent has on a high weight on the value taste. In the first instance, the agent rejects the coffee as not upholding their higher standards on the value; in the second instance, the agent is satisfied and drinks the coffee.
\texttt{propBaseClean} represents an idealised view of an agent's assessment of a set of propositions, where all propositions must pass an agent's highest weighting; presentation of other, less stringent variations remain for future work.

Any set of \texttt{propBaseClean} may contain \textit{objectively incompatible} propositions. Note that such propositions can pass the filter defined on \texttt{propBaseClean}, even if such pairs appear in \texttt{incompProp}. 
Moreover, we can see from \texttt{propBaseClean} that two agents may each accept the same proposition, yet for different settings of values and weights.

We assume  \texttt{propBaseClean} represents \textit{static} (all at once) and \textit{private} (inaccessible to others) associations of value-weights to expressions by an agent.
Note that, we do not analyse whether the propositions in \texttt{propBaseClean} are true or believed to be true; the set presents propositions which the agent can accept in the light of his/her value profile. 
The key point of creation of \texttt{propBaseClean} is to distinguish propositions which are coherent with the agent's value profile. By the same token, the complement to \texttt{propBaseClean$_{agent_h}$} reflects all those expressions which are incompatible with the Agent's values and weights.

\begin{itemize}
    \item [] \texttt{$\overline{propBaseClean}$}$_{agent_\alpha}$ =\\
\{$p_{\beta}$ $\mid$ $ p_{\beta} \not \in$ \texttt{propBaseClean$_{agent_h}$} \}
\end{itemize}

Consider the various ways propositions may appear in intersecting or complementing sets of two Agents' \texttt{propBaseClean}. For intersecting: the Agents have the same value-weight profile and same value-weights on same prop; same value-weight profile and different value-weights on prop, but not sufficient to block; different value-weight profile and same value-weight on prop, but not sufficient to block; different both, but not sufficient to block. Where they have the same props, neither the value-weight profile nor value-weight on prop is sufficient to discriminate. For complementing: same value-weight profile and different value-weights on prop, and sufficient to block; different value-weight profile and same value-weight on prop, and sufficient to block; different both and sufficient to block. In other words, differences in sets of props arise where value-weight profiles or value-weights on props are sufficient to discriminate. Broadly speaking, where a proposition appears in the intersection of the sets of \texttt{propBaseClean} of two Agents, we can say the agents agree on that proposition in one sense or another, while where the proposition is in complementary distribution (in one set, but not the other), we say there is some sense of disagreement. Note that there may be different justifications for the agreement or disagreement as well as different extents of such justification, e.g., greater or lesser difference in weights associated with the value. Relatedly, two agents can have the same denotations for their respective \texttt{propBaseClean}, yet different value-weight profiles or different value-weights on the same prop.

\begin{example}\label{ex:valueWeight}[Creating propBaseClean]

\noindent To illustrate \texttt{propBaseClean}s, suppose
an agent (agent$_A$), 4 propositions $\{p_{1},$\\$ p_{2}, p_{3}, p_{4} \}$, and 2 values: $V = \{v_{Q}, v_{P}\}$, e.g., \textit{privacy} or \textit{law enforcement}, where\\

\noindent \texttt{AgentValueToWeight}$(agent_A, value_Q)=1$\\
\texttt{AgentValueToWeight}$(agent_A, value_P)=2$\\
\noindent Agent$_A$ has high requirements concerning value $Q$, and lower requirements concerning value $P$

\noindent For \texttt{AgentValuePropWeight}, Table \ref{table:1} is a tabular form for instances of agent, propositions, values, and weights. The propositions that are in the \texttt{propBaseClean} are indicated in bold.  

\begin{table}
\caption{AgentValuePropWeigh for $agent_{A}$}
\label{table:1}
\begin{tabular}{|l|l|l|l|} 
\hline
\textbf{Agents} & \textbf{Propositions} & \textbf{Values} & \textbf{Weights} \\ [0.5ex] 
\hline\hline
agent$_{A}$ & p$_{1}$ & value$_{Q}$ & 2\\
\hline
agent$_{A}$ & $p_{1}$ & value$_{P}$ & 1\\
\hline
\textbf{agent$_{A}$} & \textbf{p$_{2}$} & \textbf{value$_{Q}$} & \textbf{2}\\
\hline
\textbf{agent$_{A}$} & \textbf{p$_{2}$} &  \textbf{value$_{P}$} & \textbf{2}\\
\hline
\textbf{agent$_{A}$} & \textbf{p$_{3}$} & \textbf{value$_{Q}$} & \textbf{3}\\
\hline
\textbf{agent$_{A}$} & \textbf{p$_{3}$} & \textbf{value$_{P}$} & \textbf{2}\\
\hline
\textbf{agent$_{A}$} & \textbf{p$_{4}$} & \textbf{value$_{Q}$} & \textbf{1}\\
\hline
\textbf{agent$_{A}$} & \textbf{p$_{4}$} & \textbf{value$_{P}$} & \textbf{3}\\
\hline
\end{tabular}
\end{table}
\end{example}

\section{State Transitions}\label{sec:STRIPS}


In order to represent our model we introduce our definitions and assumptions: 

\begin{mydef}
    Let $S$ be a set of states. Each state $s_\beta \in S$ is a consistent set of propositions. A set of propositions will be consistent if: $\forall p_\gamma, p_\delta \in s_\beta:\\ <p_\gamma, p_\delta> \not \in IncompProp$
\end{mydef}


Our basic intuition is that an agent aims to execute those actions (state transitions) relative to a precondition state which make the postcondition state more compatible with their values; that is, actions remove propositions found in the propBaseClean complement from the precondition and introduce propositions associated with propBaseClean into the postcondition. A simple example is:

\begin{example}\label{ex:stateChange1}[State Change 1]
\noindent AgentA propBaseClean: \{p1, p2\}\\
Initial-state: \{p1,p3\}\\
Action a1: $<\{p1\}, \{p1,p2\}>$, 
Action a2: $<\{p1\}, \{p2, p3\}>$\\
Action a1 is compatible with AgentA's props (relative to their values), while Action a2 is not. 
\end{example}

To model of ethical decision making into a state transition system, we use \url{https://en.wikipedia.org/wiki/Stanford_Research_Institute_Problem_Solver}{STRIPS} \cite{fikes1971}, which is a simple, familiar, relatively neutral model of a transition system (for alternatives see Section \ref{sec:RelatedWork}). STRIPS is used in planning actions from some initial state through intermediate states to a final goal state. For our purposes here, we don't introduce the planning algorithm or a choice mechanism between two equally executable actions, which are left for future work. 

\subsection{STRIPS}

Mathematically, a STRIPS instance is a quadruple $\langle P,O,I,G\rangle$, in which each component has the following meaning:
\begin{itemize}
\item $P$ is a set of conditions (i.e., propositional variables\footnote{We can take a specification of conditions as a state.});
\item $O$ is a set of operators (i.e., actions); each operator is itself a quadruple $\langle x ,y ,z ,t \rangle$, each element being a set of conditions. These four sets specify, in order, which conditions must be true for the action to be executable, which ones must be false, which ones are made true by the action and which ones are made false; 
\item $I$ is the initial state, given as the set of conditions $P$ that are initially true (all others are assumed false);
\item $G$ is the specification of the goal state; this is given as a pair $\langle N,M\rangle$, which specify which conditions are true and false, respectively, in order for a state to be considered a goal state (N represents a set of propositions that must be true, M a set of propositions that must be false).
\end{itemize}

A state is represented by the set of conditions that are true in it. Transitions between states are modeled by a transition function, which is a function mapping states into new states that result from the execution of actions. Since states are represented by sets of conditions, the transition function relative to the STRIPS instance $\langle P,O,I,G\rangle$ is a function:
\begin{itemize}
    \item [] $succ :2^{P}\times O\rightarrow 2^{P}$
\end{itemize}
In order to make a plan, the function $succ$ can be recursively extended to sequences of actions. Detailed description of this mechanism can be found in \cite{fikes1971}. If $O_1$ and $O_2$ are operators, then by $[O_1, O_2]$ we denote plan containing performing $O_1$ followed by $O_2$. 

For this paper and given space constraints, we assume:
\begin{itemize}
    \item Inertia holds between state changes.
    \item Some approach to the \url{https://en.wikipedia.org/wiki/Frame_problem}{Frame Axiom}.
    \item States are objectively consistent.
    \item An action's preconditions and postconditions are objectively consistent.
    \item We model a small sample of actions rather than all possible actions.
    \item We do not here assume that it is necessary for goals to be attained by executions of actions, e.g., relating to \textit{safety} and \textit{liveness}.
    \item In the definition of the goal G, N is what the agent ``wants'' and M is what the agent doesn't ``want'', i.e., $propBaseClean_{agent_\alpha}$ and $\overline{propBaseClean_{agent_\alpha}}$.\footnote{Whether G represents the agent's \textit{ideal}, everything they ``want'' and nothing they don't ``want'', or some \textit{subideal}, somethings they ``want'' with something they don't ``want'', is a modeling issue for future development.}
\end{itemize}


\subsection{STRIPS$_{VFR}$ - Relative to an Agent's propBaseClean}

However, we want to relate the actions that an agent executes to that agent's propBaseClean to imply the compatibility with that agent's value profile.

We assume the goal state G, $<N,M>$, where $N$ is consistent and relative to $propBaseClean_{agent_\alpha}$ (also see the footnote 8).

\begin{mydef}[Goals Relative to Agent's propBaseClean]\label{def:goalsAgents}
Where G$_{agent_\alpha}$ of $\langle P,O_{agent_\alpha},I,G_{agent_\alpha}\rangle$ is\\ $<N_{agent_\alpha},M_{agent_\alpha}>$:\\
$N_{agent_\alpha}$~$\in$ $2^{propBaseClean_{Agent_\alpha}}$, where\\
$\forall \gamma, \delta \in$ $N_{Agent_\alpha}$:\\ $<\gamma, \delta> \not \in IncompProp$.
\end{mydef}

Note that a \textit{goal} is, in STRIPS, simply a proposition that holds of a goal state. In VFR, a value is realised with respect to propositions, i.e., those in an Agent's \texttt{PropBaseClean}. In this sense, a goal connects with an Agent's values.

We define those actions which lead to the introduction of a proposition relative to $propBaseClean_{agent_\alpha}$:

\begin{mydef}[Agent's Actions Introducing Propositions]\label{def:actions1}
We revise $O$ of $\langle P,O,I,G\rangle$ to relativise the actions to the agent -- $\langle P,O_{agent_\alpha},I,G_{agent_\alpha}\rangle$.\\
$O_{agent_\alpha}$ is a quadruple $\langle x,y,z_{agent_\alpha},t \rangle$, where $x,y$ and $t$ are as in $O$, where:\\
$z_{agent_\alpha}$ $\subseteq$ $propBaseClean_{agent_\alpha}$.
\end{mydef}
The revised definition of actions means that the actions an agent takes has to result in a postcondition where propositions hold that are compatible with the agent's values. It does not execute actions which introduce propositions incompatible with the agent's values. However, it also does not necessarily eliminate from the precondition those propositions which are incompatible with the agent's values and which my be inherited by inertia.

The transition function is also modified $\langle P,O_{agent_\alpha},I,G_{agent_\alpha}\rangle$ is a function:
\begin{mydef}[Revised Transition Function]\label{def:revTransition}
$succ :2^{P}\times O_{agent_\alpha} \rightarrow 2^{P}$
\end{mydef}

The above modification of the STRIPS framework allows the agent to take only the actions which lead to those consequences that are wanted from the perspective of that agent, i.e. consequences which pass the value profile filter and are in $propBaseClean_{agent_\alpha}$. In other words, our model can be understood as a kind of filter which excludes from the available plans those which do not lead to consequences that satisfy the Agent's values and includes those which do. 
Note also that this model does not preserve the system from \textit{inheriting} (by inertia) bad propositions from previous states.
\begin{example}\label{ex:actions2}
    Suppose a set of propositions $\{p_1, p_2, p_3, p_4\}$ and a \\propBaseClean$_{agent_A} = \{p_2, p_3, p_4 \}$.
    Suppose initial state $\{p_1, p_2\}$ (note that initial state is not in agent's propBaseClean)
Set $O_{aget_A}$ contains following operators:
\begin{itemize}
    \item[01:] $\langle \{p_1, p_2\},\{\}, \{p_3\},\{p_1\}\rangle$
    \item[02:] $\langle \{p_2\},\{\}, \{p_3\},\{\}\rangle$
    \item[03:] $\langle \{p_3\},\{\}, \{p_4\},\{\}\rangle$
\end{itemize}
Goal of the agent is: $G_{agent_A} = \langle \{p_4\}, \{\}\rangle$
In standard STRIPS there are two available plans: 
\begin{itemize}
    \item $[O1, O3]$ which, in final state, contains: $s_{final} = \{p_2, p_3, p_4\}$
    \item $[O2, O3]$ which, in final state, contains: $s_{final} = \{p_1, p_2, p_3, p_4\}$\footnote{Given inertia, neither $O2$ nor $O3$ remove $p_1$.}
\end{itemize}
Since the second plan (following the inertia of state changes) contains proposition $p_1$ which is not in the propBaseClean of the agent, so leads to a subideal world from the perspective of the Agent.\footnote{The revised  definition of O and example can be further refined.}
If we introduce another action: 
\begin{itemize}
    \item[03':] $\langle \{p_3\},\{\}, \{p_1,p_4\},\{\}\rangle$
\end{itemize}
The number of available plans allowing for reaching the goal for standard STRIPS will increase. For 
STRIPS$_{VFR}$, however, it will remain the same, because operator O3' leads to ``immoral'' results (it brings about $p_1$ which is outside of \texttt{propBaseClean} and not acceptable for the agent) and cannot be used. 
\end{example}

If we want actions relative to an Agent to incrementally remove impurities, then we need to identify those propositions of the precondition which are unacceptable to the Agent, i.e., elements that in $\overline{propBaseClean_{agent_\alpha}}$, then to remove them from the postcondition state. In this way, the Agent incrementally acts to make the world more compatible with their $propBaseClean_{agent_\alpha}$.

\begin{mydef}[Agent's Actions Removing Bad Things]\label{def:agentremoving}
$O^+_{agent_\alpha}$ is a set of quadruples $\langle x,y,z_{agent_\alpha},t^+ \rangle$, where:\\
$\forall$ $\alpha$: [[$\alpha$ $\in$ $x$ $\wedge$ $\alpha$ $\in$ $\overline{propBaseClean_{agent_\alpha}}$] $\rightarrow$ $\alpha$ $\in$ $t^+$].
\end{mydef}

This is a filter on actions, those which meet the conditions. It says that there can be actions that clean with respect to $propBaseClean_{agent_\alpha}$, but there may be actions that are not defined by the Agent and so an undesireable proposition can remain and be inherited. That is, it may only be partial in terms of cleaning `power', but it is not necessarily total.\footnote{Alternative proposals are possible, but left as future work.} In Example \ref{ex:actions2}, Action 01 `positively' satisfies Definition \ref{def:agentremoving}, while Actions 02 and 03 vacuously satisfy the definition. An action such as Action 01' $\langle\{p_1, p_2\},\{\}, \{p_3\},\{\}\rangle$ does not satisfy the definition.

\section{Implementation}

In order to verify our model, we prepared its experimental implementation in PROLOG language. Our program uses standard PROLOG reasoning mechanism, below we present fragments of the code and discuss some key predicates. 

We modelled the examples (1,3) presented in the paper. In particular, the mechanism of propBaseClean creation as well the ethical plan selection. The listing below presents a fragment of the propBaseClean creation:
\begin{verbatim}
proposition(p1).
proposition(p2).
proposition(p3).
proposition(p4).
agentValue(valueP,2).
agentValue(valueQ,1).
agentValueProp(p1,valueP,1).
agentValueProp(p1,valueQ,2).
agentValueProp(p2,valueP,2).
agentValueProp(p2,valueQ,2).
agentValueProp(p3,valueP,2).
agentValueProp(p3,valueQ,3).
agentValueProp(p4,valueP,3).
agentValueProp(p4,valueQ,1).
notpassValue(X,Val):- agentValueProp(X,Val,Wp),
     agentValue(Val,Wa), >(Wa,Wp).
propBaseClean(X):- proposition(X), \+ notpassValue(X,\_).
    ...
\end{verbatim}
Predicate \texttt{notpassValue(X,Val)} defines which propositions do not pass the value test, where \texttt{AgentValueToWeight} for a given value's weight (variable \texttt{Wa}) is higher than the weight assigned to the same value with respect to a proposition (variable \texttt{Wp}). Predicate \texttt{propBaseClean(X)} creates a list of propositions which pass the test (fails not passing test).

The fragment mechanism of plan creation is presented below:

\begin{verbatim}
oper([p1,p2],[],[p3],[p1]).
oper([p2],[],[p3],[]).
...
goal([p4],[]).

ethical(L):-findall(X,propBaseClean(X),L).

satisfied(X):- goal(Z,T), in(Z,X), \+ commonel(X,T).

ethMove(X,Y):-ethOper(Xp,Xn,Zp,Zn), in(Xp,X), \+ commonel(X,
    Xn), subtract(X,Zn,Xdel), union(Xdel,Zp,Y), \+in(Y,X).
ethOper(Xp,Xn,Zp,Zn):- oper(Xp,Xn,Zp,Zn), ethical(E), 
    in(Zp,E).

ethPlan(X,[X]):- satisfied(X).
ethPlan(X,Y):- ethMove(X,Z), ethPlan(Z,Y).
\end{verbatim}
Predicate \texttt{ethPlan(X,Y)} (implementing Revised Transition Function, see def. \ref{def:revTransition}) represents the recursive mechanism of the creation of a plan, predicate \\ \texttt{ethOper(Xp,Xn,Zp,Zn)} (implementing Agent’s Actions Introducing Propositions, see def. \ref{def:actions1}) selects operators the consequences of which are in \texttt{propBaseCLean} (variable Xp represents conditions which must be true, Xn represents propositions which must be false, Zp represents propositions which becomes true, Zn propositions which becomes false). Subsidiary predicates \texttt{in(X,Y)} and \\ \texttt{commonel(X,Y)} define whether propositions X are also in Y or whether two lists of propositions have common elements. Predicate \texttt{satisfied(X)} defines which states fulfill the goal. Predicate \texttt{ethMove(X,Y)} defines move from one state to another, taking into consideration only these operators which lead to acceptable (w.r.t. \texttt{propBaseClean}) consequences\footnote{Full version of our experiment and definitions of other predicates can be found here: \url{https://github.com/ZurekTom/PrologStrips/tree/main}}. 

If we provide a query \texttt{ethPlan([p1,p2],[p2,p3,p4])} then the system will check whether there is an ethical plan leading from [p1,p2] to [p2,p3,p4] (the answer is \texttt{true}). Note that the result list should contain a goal (in our case p4). If we ask \texttt{ethPlan([p1,p2],X)}, then the system will return two possible ethical plans leading to a declared goal. Note that if add another action, e.g., \texttt{oper([p3],[],[p1,p4],[])}, the result will be the same (there will be no more ethical plans) because this operator leads to an ``unethical'' result (it brings about p1 which does not pass the \texttt{propBaseClean} test). 

\section{Related Work and Discussion}\label{sec:RelatedWork}
\paragraph{Related Work} There are other ways to represent and reason with actions, but our main objective is to formalise actions given an agent's values. We used the well known, simple mechanism of STRIPS, leaving the more advanced approaches to actions -- situation calculus, event calculus, BDI, Dynamic Logic, Action Logics, AATS, and others -- for a future analysis. Below we discuss a selection of approaches to modeling consequentialist ethics:

Most of the existing approaches to  ethical decision making use a kind of ordering between values. For example,
\cite{DBLP:journals/ai/WinikoffSDD21} augment a Belief-Desire-Intention framework with values, where valuings indicate preferences amongst outcomes. The treatment of values is somewhat similar to our approach in that they are abstract properties associated with propositions, but they introduce conditional orderings over effects on values caused by decision options; these are not necessary in our model and would seem to introduce a high degree of complexity.


Values appear in discussions about actions in multi-agent systems with Action-based Alternating Transition Systems with values (AATS+V) 
\cite{DBLP:journals/flap/AtkinsonB21}. Essentially, an AATS+V expresses actions as transition functions from state to state, where states are sets of propositions. Values are a set of abstract objects. A valuation function describes whether a state transition promotes or demotes the value; in a sense, it is a preference over actions. However, the values relate to states as wholes, i.e., a set of propositions. Thus, it is unclear just what it is about the states, i.e., particular propositions, such that the transition promotes or demotes the value.  STRIPS$_{VFR}$ is more specific, as the Agent executes actions which bring about states containing propositions that are more compatible with its values. In AATS+V actions are made in order to promote, i.e. increase the levels of satisfaction of values), without considering that a given value can be already promoted to the satisfactory level (drinking one beer can promote happiness, but drinking 8 beers not, as it will result in hangover). In other words, AATS+V does not allow for expressing whether a given value reached a necessary level of satisfaction.  An extended version of AATS+V \cite{DBLP:journals/flap/AtkinsonB21} introduces the mechanisms called `maximisers' and `satisficers', but these mechanisms are very complex and it is problematic how to represent that a given value does not have to be promoted to a higher level than it is necessary.   

Preferences are considered in AI across topic areas where choices are made between objects and preference is a comparative ordering of the objects \cite{DBLP:journals/ai/DomshlakHKP11}. In our proposal, there is no direct comparison between objects (propositions) and an ordering over them. Rather, values filter propositions according to the value profile of the Agent; essentially, this identifies the belief set that the Agent uses for reasoning. 


Our model shares some similarities with two value-based decision making models: \cite{zurekgvr} and later \cite{heidari2020}, which were constructed on the basis of similar assumptions. Firstly, both models are rooted in Schwartz value theory \cite{Schwartz2012} (\cite{heidari2020} focuses strictly on the list of values introduced in Schwartz, et al., while \cite{zurekgvr} has more abstract character without specifying concrete values). Secondly, both models assume that the thresholds on the levels of the satisfaction of values can function as a motivation of the decisions made by an agent. The differences between these models are in the details of formal machinery used to model the analysed phenomena and in the burden of analysis: \cite{zurekgvr} focuses on the decision making process, while \cite{heidari2020} on the relations between values. The most similar to our approach is \cite{zurekstachura2021}, which is based on a formal model from \cite{zurekgvr}, where the \cite{zurekgvr} is discussed from the point of view of various ethical theories. Our approach extends the above models by introducing a simple, but universal reasoning mechanism and by introducing refined description of the states (state is defined not by one proposition, but by a set of propositions), which allows for a more accurate representation of the actual states. 

Most existing approaches to consequentialist ethics in knowledge-based systems focus on the utilitarian version of consequentialism (for example \cite{anderson2005}) where the authors focus on maximisation of a utility function. Our approach is different because we do not aim at maximizing utility, but rather introduce a mechanism to exclude ``immoral'' consequences from a plan.

Recently, a number of approaches to consequentialist ethics in reinforcement learning systems have been introduced (for example, \cite{rodriguezsoto2021}). Although they are out of the scope of this paper, it is worth considering whether our approach could be useful in embedding ethics into reward function.




\paragraph{Discussion} In this paper, we introduce a new mechanism wherein an Agent's actions take into consideration the Agent's value profile. We modify a STRIPS planner (STRIPS$_{VFR}$) as an example of an action (state transition) system, but our model can be adapted (after necessary modifications) to other action formalisms and mechanisms.

Our research based on the observation that the ethical evaluation of decision options is made on the basis of two individual scales: a value profile which describes the individual motivations of an Agent concerning the Agent's values (which in our paper is represented by a set of thresholds); and a personal evaluation of the consequences of decisions (which in our work is represented as a weights of values assigned to particular propositions). The decision about which plan to choose is made on the basis of the suitability of the evaluation of the consequences of the decision with respect to the agent's value profile. 

Such an approach has a number of consequences: firstly it individualises the value-based evaluation of decisions, rather homogenises them in a single universal value hierarchy. Such an approach represents real life situations, because not every Agent shares the same needs, point of view, or priorities. Secondly, our model gives a possibility for implementing various approaches to deal with undesired propositions: from ignoring them (a ``liberal'' option, see def. \ref{def:actions1}) to removing all non-acceptable proposition (``totalitarian'' option). Moreover, the approach relates to conceptions of multi-agent systems where the behaviour of agents, whether individually or as groups, is the result of interactions amongst individuals and their preferences (bottom-up) rather than globally (top-down). Voting is an example.

Our approach can be seen as corresponding to a number of consequentialist claims. The focus on consequences represented by states of affairs contributes to an essentially consequentialist outlook on the moral assessment of possible actions. The Agent filters out propositions, which ensures that the states of affairs are good enough in moral terms; this accounts for the satisficing variant of consequentialism. An individualized value profile of the agent corresponds to the preferential nature of the ethical approach we have adopted.

Our model allows for a multi-agent approach to behaviour which reflects value choices, e.g., introducing more than one agent and relativisation of STRIPS to different agents: agents have their own ethical principles expressed by the VFR mechanism and their propBaseCleans, individualised set of operators, and their own goals.
On the basis of that we can discuss the possibility of negotiation and cooperation amongst agents.
The basic intuition is, that two agent can cooperate if any proposition in the intermediate or final states of those agents will be in propBaseCleans of both agents. This topic requires more detailed discussion.

In the proposal, an agent executes those actions wherein the consequences are compatible with that agent's value profile. In this sense, the proposal is a constraint on the actions which are available to an agent to execute; that is, the agent can only execute actions compatible with their values. In this, it is an abstraction and simplification. In particular, the proposal does not introduce issues related to the selection of actions or plans, wherein an agent might execute a ``suboptimal'' action not wholly compatible with their values. In such an instance, one might consider whether the values an agent aims to achieve can justify ``suboptimal'' actions. Such matters raise a range of issues about moral and ethical reasoning, which are for future work.

Finally, we aim to develop a \textit{family} of approaches of different settings of parameters (goals, action definitions, relations of actions to goals) that have different computational properties. Such approaches would be associated with alternative theories of values in philosophy. By this we could enable computational exploration of various forms of ethical theories.


%
%
%
 \bibliographystyle{splncs04}
 \bibliography{bibliography}
%




\end{document}